\documentclass[aps,nofootinbib,showpacs,showkeys,preprint
,tightenlines ] {revtex4}

\usepackage{epsf,epsfig,subfigure,graphicx,amsmath,amssymb}
\usepackage{color}

\newcommand{\dis}[1]{\begin{equation}\begin{split}#1\end{split}\end{equation}}
\newcommand{\be}{\begin{equation}}
\newcommand{\ee}{\end{equation}}

\newcommand{\eq}[1]{Eq.~(\ref{#1})}

\newcommand{\bfrac}[2]{{\left(\frac{#1}{#2} \right)  }}
\newcommand{\Mp}{M_P}

\newcommand{\gev}{\,\textrm{GeV}}

\newcommand{\mpc}{\,\textrm{Mpc}}

\def\bea{\begin{eqnarray}}
\def\eea{\end{eqnarray}}

\begin{document}

 \begin{flushright}
{\tt  
}
\end{flushright}

\title{\Large\bf Large tensor spectrum of BICEP2
\\in the natural supersymmetric hybrid model}

\author{Ki-Young Choi$^{(a)}$\footnote{email: kiyoungchoi@kasi.re.kr}
and Bumseok Kyae$^{(b)}$\footnote{email: bkyae@pusan.ac.kr} }
\affiliation{$^{(a)}$ Korea Astronomy and Space Science Institute, Daejon 305-348, Republic of Korea\\
$^{(b)}$ Department of Physics, Pusan National University, Busan
609-735, Republic of Korea}


\begin{abstract}
The large tensor spectrum recently observed by the BICEP2 Collaboration requires a super-Planckian field variation of the inflaton in the single-field inflationary scenario. 
The required slow-roll parameter $\epsilon\approx 0.01$ would restrict  
the e-folding number to around 7 in (sub-)Planckian inflationary models.  
To overcome such problems, we consider a two-field scenario based on the natural assisted supersymmetric (SUSY) hybrid model (``natural SUSY hybrid inflation'' \cite{NHinf}), 
which combines the SUSY hybrid and the natural inflation models. 
The axionic inflaton field from the natural inflation sector can admit the right values for the tensor spectrum as well as a spectral index of $0.96$ with a decay constant smaller than the Planck scale, $f\lesssim M_P$. 
On the other hand, the vacuum energy of $2\times 10^{16} \gev$ with 50 e-folds 
is provided by the inflaton coming from the SUSY hybrid sector, avoiding the eta problem. 
These are achieved by introducing both the U(1)$_R$ and a shift symmetry, and employing the minimal K$\ddot{\rm a}$hler potential. 
\end{abstract}


 \maketitle


\section{Introduction}

Cosmological inflation not only resolves the problems in the standard big bang cosmology such as the homogeneity and flatness problems, but also explains the cosmological perturbations in matter density and spatial curvature \cite{review}.  
Those could naturally arise from the vacuum fluctuations of light scalar field(s) during inflation and be promoted to classical fluctuations around the time of the horizon exit. 
Indeed, the primordial power spectrum generated by an inflaton \cite{InflationFluctuation} turned out to be quite consistent with the cosmic microwave background (CMB) observations \cite{COBE,Komatsu:2010fb,Ade:2013zuv}. 
At the present, the inflationary paradigm seems to be strongly supported by various  cosmological observations. 

Recently, the BICEP2 Collaboration reported their observation on the B mode  
of the primordial gravitational wave \cite{Ade:2014xna}. 
Their measurement indicates a large tensor spectrum or a large tensor-to-scalar ratio in the power spectrum \cite{Ade:2014xna}: 
\dis{
r=0.2^{+0.07}_{-0.05}\quad ({\rm or}~~ 0.16^{+0.06}_{-0.05} ~), 
}
(after foreground subtraction with the best dust model).  
The tensor spectrum is evaluated on the scale, 
$x_{\rm ls}/100 \lesssim k^{-1} \lesssim x_{\rm ls}$ with $x_{\rm ls} =  14,000 \mpc$, which corresponds to 
a change in the e-folding number, $\Delta N \sim 4$. 
For smaller scales, however, the tensor spectrum by the primordial gravitational wave 
is suppressed. Moreover, it is not constrained by observation anymore.
Such a large tensor-to-scalar ratio for the large scale implies relatively large values of the slow-roll parameter $\epsilon$ and the vacuum energy during that period of inflation: 
\dis{ \label{imply}
\epsilon^*\approx 0.01 \quad {\rm and}\quad 
V^{1/4}\approx 2.08\times 10^{16} \gev .
} 

In the slow-roll regime, the large $r$ requires a {\it super}-Planckian field variation for the inflaton in single-field inflationary models \cite{Lyth:1996im,lyth2}:
\dis{ \label{lythBD}
\frac{\Delta \varphi}{\Mp} \gtrsim \mathcal{O} (1) \times \bfrac{r}{0.1}^{1/2},
}
where $M_P$ denotes the reduced Planck scale ($\approx 2.4\times 10^{18} \gev$).
However, a super-Planckian field variation might imply the breakdown of an effective field theory description on inflation. 
On the other hand, the problem of {\it sub}-Planckian inflation is that such a relatively large $\epsilon \approx 0.01$ yields a too small e-folding number:  
\dis{ \label{efoldBD}
\Delta N \approx \frac{1}{\Mp} \int \frac{d\varphi}{\sqrt{2\epsilon}} \approx 7 ~\bfrac{\Delta \varphi}{\Mp} \sqrt{\frac{0.01}{\epsilon}} .
}
Thus, only  $\Delta N \sim 7$ is maximally obtained for $\Delta \varphi\sim M_P$. 
In order to get a large enough e-folding number, hence, either the field value must be super-Planckian or the slow-roll parameter $\epsilon$ should somehow be made to rapidly decrease (after about 7 e-folds \cite{Choi:2014aca}).\footnote{For the possibility of a large tensor-to-scalar ratio with single-field inflation models, see Refs.~\cite{Choudhury:2014kma,Antusch:2014cpa}.} 
For super-Planckian field values, however, it is necessary to invoke a symmetry to avoid the disastrous higher-order terms~\cite{Freese:1990rb,Kim:2014dba}. For a too rapidly decreasing $\epsilon$, the power spectrum may violate the observational constraint.

To be consistent with the observation of the CMB,   
the power spectrum should be maintained as almost a constant within the observable scales in the CMB, 
which is within the range of $10 \mpc \lesssim k^{-1} \lesssim x_{\rm ls}$,  corresponding to to $\Delta N\sim 7$ \cite{Ade:2013zuv}.  
For smaller scales, the power spectrum is constrained only to be smaller than around $10^{-2}$  by  the argument of the missing primordial black hole \cite{Khlopov:1980mg}, or  $10^{-4}$ by acoustic damping \cite{Chluba:2012we,Khatri:2013dha}, and especially less than 0.007 for $10^{-5}  \mpc \lesssim k^{-1} \lesssim 10^{-4}  \mpc $ from big bang nucleosynthesis (BBN)~\cite{Jeong:2014gna}. 
There is also a stronger constraint, ${\cal P}_\zeta \lesssim 10^{-6}$,  for a specific type of dark matter using the nonobservation of the ultracompact mini halo \cite{Bringmann:2011ut}.\footnote{The improved observation on the spectral distortions on the CMB will constrain the power spectrum more strongly in the large scales corresponding to  $\Delta N\sim 17$ from $x_{\rm ls}$ \cite{Chluba:2012we,Khatri:2013dha}.}
Actually, it is quite hard to accommodate all the above stringent constraints within the single-field inflationary framework. 

The inflationary scenario is, indeed, based on the quantum theory of scalar fields. 
For successful inflation, the inflaton mass is required to be much lighter than the Hubble scale during inflation. 
As seen in the Higgs boson and the gauge hierarchy problem in elementary particle physics, 
however, 
it is highly nontrivial to keep a small enough inflaton mass against quantum corrections.
Actually, only a few ways to get a light scalar are known in quantum field theory: i.e., 
by introducing {\bf (i)} the supersymmetry (SUSY), {\bf (ii)} a global U(1) symmetry, or 
{\bf (iii)} a strong dynamics. 
We will discuss here only the first two possibilities for a small inflaton mass. 

SUSY is an excellent symmetry that can protect a small scalar mass against quantum corrections.  
However, the problem in SUSY inflationary models is that 
SUSY must be broken due to positive vacuum energy of the Universe during inflation, 
even if it was introduced: 
positive vacuum energy inflating the Universe can induce a Hubble scale inflaton mass in supergravity (SUGRA),  
violating a slow-roll condition, $\eta\sim {\cal O}(1)$.  
It is called the ``eta ($\eta$) problem'' in SUGRA inflation models. 
It is a quite generic problem requiring an elaborate model construction to overcome.     

In SUSY hybrid inflation or ``$F$-term inflation'' \cite{FtermInf2,FtermInf}, 
fortunately the Hubble induced mass term is accidentally canceled out with the minimal K$\ddot{\rm a}$hler potential and the Polonyi-type superpotential during inflation. 
The specific form of the superpotential can be guaranteed by the introduced U(1)$_R$ symmetry. 
In this class of models, SUSY-breaking positive vacuum energy 
generates a logarithmic quantum correction to the constant scalar  potential, which can draw the inflaton to the true minimum, 
triggering reheating of the Universe by the waterfall fields. 
The waterfall fields develop nonzero vacuum expectation values (VEVs) at their true minima, which 
can be determined with CMB anisotropy \cite{FtermInf}. 
They turn out to be tantalizingly close to the scale of the grand unified theory (GUT), ($\approx 2\times 10^{16} \gev$). 
Due to this, the waterfall fields can be regarded as GUT-breaking Higgs bosons in this class of models \cite{3221,422,FlippedSU(5),SO(10)}.

In the (original) SUSY hybrid inflation model, a red-tilted power spectrum \cite{FtermInf} around 
\begin{equation}
n_\zeta \approx 1+2\eta\approx 1-\frac{1}{N_e}\approx 0.98
\end{equation}
is predicted for $N_e=50$ -- 60 e-folds. 
It is too large compared to the present bound on the spectral index. 
On the contrary, the tensor spectrum is too small to detect, 
$r\lesssim 0.03$ in this class of models  \cite{Shafi:2010jr,Rehman:2010wm,Okada:2011en,Civiletti:2011qg}.
Basically, the required slow-roll parameter, $\epsilon\approx 0.01$, is hard to get in SUSY hybrid models.

In the ``natural inflation'' model \cite{Freese:1990rb}, the inflaton is looked upon as a pseudo-Nambu-Goldstone boson introduced from the spontaneous breaking mechanism of an anomalous global U(1) symmetry, U(1)$_{\rm PQ}$. 
By instanton effects, which break U(1)$_{\rm PQ}$ to a shift symmetry, 
a sinusoidal-type inflaton potential can be generated in this model.  
Because of a remaining shift symmetry,
the inflaton field does not appear in the K$\ddot{\rm a}$hler potential in the SUSY version of the natural inflation model \cite{NaturalChaotic}. 
As a result, the unwanted Hubble scale inflaton mass term is not induced in the SUGRA potential during inflation. 

For a small enough slow-roll parameter $\eta$, however, the U(1)$_{\rm PQ}$ breaking scale or the ``axion decay constant'' $f$ must be larger than the Planck scale, 
\begin{equation}
f \gtrsim 3 M_P ~. 
\end{equation}
This implies that U(1)$_{\rm PQ}$ should be valid above the Planck scale.
However, such a U(1)$_{\rm PQ}$ is not natural, because quantum gravity effects are known to break all continuous global symmetries, including U(1)$_{\rm PQ}$.
One possible way to obtain an effectively large $f$ from a sub-Planckian Peccei-Quinn scale 
is to employ multiple axionic inflaton fields \cite{Kim:2004rp,doubleAxion}.

Interestingly enough,  the ``natural'' assisted SUSY hybrid inflation model or ``natural hybrid (NH) inflation'' \cite{NHinf}, which combines the SUSY hybrid and the natural inflation models,  
can cure the problems of the SUSY hybrid and natural inflationary models by interplay of the two inflatons: 
it can yield the desired value $n_\zeta\approx 0.96$ and realize $f\ll M_P$. 
The NH inflation introduces a shift and U(1)$_R$ symmetries, and it provides two light inflaton fields: 
one is the SUSY hybrid inflaton, and the other is the axionic one.
Their Hubble induced masses are not generated if the K$\ddot{\rm a}$hler potential is of the minimal form \cite{NHinf}. 
Thus, the $\eta$ problem can be avoided. 
The smallness of the higher-order terms (particularly the quartic coefficient) in the K$\ddot{\rm a}$hler potential, however, needs to be justified with a UV quantum gravity theory. 
Since just a shift symmetry rather than U(1)$_{\rm PQ}$ is employed, the decay constant $f$ here does not have to be associated with a spontaneous symmetry-breaking mechanism.


In this paper, we will consider a two-field inflationary scenario \cite{Choi:2012hea} by the NH inflation model to account for the observed large tensor spectrum: 
we will attempt to show that the NH inflation model achieves the large tensor-to-scalar ratio $0.16$  
by interplay of the two inflaton fields. 
A large enough e-folding number during inflation to resolve the homogeneity and flatness problems can be obtained 
with large vacuum energy before the waterfall field destabilizes the vacuum. 
It is  mainly driven by the inflaton coming from the SUSY hybrid model. 
On the other hand, the cosmological observables associated with quantum fluctuations of the inflaton 
are mainly provided by the inflaton of the natural inflation.  

During the first seven e-folds 
after the comoving scale of the last scattering exits the horizon, 
inflation is driven by the two inflaton fields, explaining the observed spectra in the CMB.
After the axionic inflaton reaches almost the minimum of its potential, however, 
the inflationary history follows that of the SUSY hybrid models 
until the waterfall fields terminate the inflation, eventually yielding about 50 e-folds \cite{doublehybrid}.  
In this model, the field values of the SUSY hybrid inflaton during inflation are sub-Planckian ($\sim 10^{16-17} \gev$), 
while the axionic inflaton moves a Planckian distance along the phase direction \cite{phase}, 
which would not cause harm.

This paper is organized as follows. In Sec. \ref{sec:model}, we set up the SUGRA model. 
In Sec. \ref{sec:Inf}, we discuss the spectrum and its indices for both scalar and tensor perturbations 
in the NH inflation model, 
which are the main results. 
We conclude in Sec. \ref{sec:conclusion}.


\section{The Natural Hybrid Inflation Model}
\label{sec:model}


In this section, we briefly review the NH inflation model \cite{NHinf}. 
We introduce the U(1)$_R$ symmetry and a shift symmetry. 
They are necessary to protect the two light inflaton masses  against the Hubble scale SUGRA corrections during inflation. 
Under the U(1)$_R$ symmetry, the superpotential $W$ and a superfield $S$ are assumed to transform in the same manner: 
$W\rightarrow e^{2i\gamma}W$ and $S\rightarrow e^{2i\gamma}S$. 
Under the shift symmetry, a superfield $T$ [$=\frac{1}{\sqrt{2}}(\phi+ia)$] is supposed to undergo the transformation as $T\rightarrow T+2\pi i f$, where the decay constant $f$ is a constant with mass dimension 1. 
Hence, only the imaginary component of $T$ transforms under the shift symmetry, 
$a\rightarrow a+2\sqrt{2}\pi f$.
$S$ and $a$ are regarded as the inflaton fields. 
The shift symmetry does not have to be embedded in a global U(1) symmetry for a small mass of $a$:  
since the SUSY has been already introduced, the inflaton $a$ can potentially be light. 
Just for avoiding the Hubble induced mass term, a shift symmetry is enough \cite{NaturalChaotic}, as will be seen below. 
We also consider the superfields of a conjugate pair, $\psi$ and $\overline{\psi}$, which are assumed to carry opposite gauge charges.  
They play the role of the waterfall fields.

The superpotential consistent with the U(1)$_R$ and the shift symmetries is written as 
\dis{ \label{globalSUSYPot}
W=\kappa S\left[M^2-\sum_n m_n^2e^{-nT/f}
-\psi\overline{\psi}\left(1+\sum_n\rho_ne^{-nT/f}\right)\right] ,
}
where $M^2$ and $m_n^2$ ($\kappa$ and $\rho_n$) are dimensionful (dimensionless) parameters. 
$n$ can be any integral number.  
We require the hierarchy between the parameters
\begin{equation} \label{hierarchy}
\frac{m_n^2}{M^2}~\ll~\frac{f^2}{M_P^2} ~\lesssim ~1 .
\end{equation}
At the minimum of the $F$- and $D$-term potentials, 
$\psi$ and $\overline{\psi}$ develop proper VEVs, 
$|\langle\psi\rangle|=|\langle\overline{\psi}\rangle|\approx M$.\footnote{In this class of models, 
a gauge symmetry is broken after the end of inflation. However, the expected cosmological problems associated with  topological defects could be avoided by introducing higher-order terms of the waterfall fields, leaving intact the salient features of the inflationary scenario \cite{422}. 
This is because such higher-order terms admit the trajectory along which the gauge symmetry is broken during inflation. 
This mechanism just effectively redefines the mass parameter $M$.}
On the other hand, for $\langle S\rangle\gtrsim M$, they are stuck to the origin, 
$\langle\psi\rangle=\langle\overline{\psi}\rangle=0$, 
because of the heavy masses generated by the nonzero VEV of $S$, and almost constant vacuum energy ($\approx \kappa^2M^4$) is induced,  
which makes inflation possible.

During inflation, thus, the waterfall fields $\psi$ and $\overline{\psi}$ are decoupled from dynamics due to their heavy masses, and so 
the superpotential and the K${\rm \ddot{a}}$hler potential are simply given by
\begin{eqnarray}
W_{\rm inf}=\kappa S\left(M^2-\sum_n m_n^2e^{-nT/f}\right)  \quad {\rm and} \quad 
K_{\rm inf}=|S|^2+\frac12(T+T^*)^2 ,
\end{eqnarray}
respectively. 
Note that the imaginary component of $T$, i.e. $a$, does not appear in the K${\rm \ddot{a}}$hler potential due to the shift symmetry.  
With the covariant derivatives in SUGRA, 
\begin{eqnarray}
&&D_SW=\frac{\partial W}{\partial S}+\frac{W}{M_P^2}\frac{\partial K}{\partial S}=\kappa M^2\left(1+\frac{|S|^2}{M_P^2}\right)\left(1-\sum_n\frac{m_n^2}{M^2}e^{-nT/f}\right) , \\
&&D_TW=\frac{\partial W}{\partial T}+\frac{W}{M_P^2}\frac{\partial K}{\partial T}\approx \kappa M^2\frac{fS}{M_P^2}\left(\frac{T+T^*}{f}
+\frac{M_P^2}{f^2}\sum_nn\frac{m_n^2}{M^2}e^{-nT/f}\right) , 
\end{eqnarray}
and the (inverse) K${\rm \ddot{a}}$hler metric,  $K^{SS^*}=1/K_{SS^*}=1$, $K^{TT^*}=1/K_{TT^*}=1$, $K^{ST^*}=K_{ST^*}=0$, etc., 
one can write down the $F$-term scalar potential:
\begin{eqnarray} \label{V_sugra}
&& V_{\rm SUGRA}=e^{K/M_P^2}\left[\left|D_SW\right|^2+\left|D_TW\right|^2-3\frac{|W|^2}{M_P^2}\right] \nonumber \\
&&\approx\kappa^2 M^4\left(1+\frac{\phi^2}{M_P^2}\right)\left[
\left|1-\sum_n\frac{m_n^2}{M^2}e^{-nT/f}\right|^2
+\frac{f^2|S|^2}{M_P^4}\left|\frac{\sqrt{2}\phi}{f}
+\frac{M_P^2}{f^2}\sum_nn\frac{m_n^2}{M^2}e^{-nT/f}\right|^2\right] \qquad \\
&& \approx \kappa^2 M^4\left[1-\left(\sum_n\frac{m_n^2}{M^2}e^{-nT/f}+{\rm h.c.}\right) 
+\frac{\delta\phi^2}{M_P^2}
\right] ,  \nonumber 
\end{eqnarray}
where we have inserted $(T+T^*)=\sqrt{2}\phi$ and dropped $|S|^4/M_P^4$ because of its smallness, and used $e^{K/M_P^2}\approx (1+\phi^2/M_P^2)(1+|S|^2/M_P^2)$. 
As seen in \eq{V_sugra}, the Hubble scale mass terms for $S$ and $a$ are not generated. 
In contrast, the real component of $T$, i.e. $\phi$ ($\equiv\langle\phi\rangle+\delta\phi$), which is invariant under the shift symmetry and so contributes to the K${\rm \ddot{a}}$hler potential, obtains the Hubble scale mass term,  $\kappa^2M^4\delta\phi^2/M_P^2=3H^2\delta\phi^2$, violating the slow-roll condition ($\eta=1$). 
Hence, it should be stabilized during inflation. Its VEV is estimated as $\langle\phi\rangle/f={\cal O}(M_P^2m_n^2/f^2M^2)\ll 1$.  
We will neglect $\langle\phi\rangle/f$.


\section{Inflation with the Large Tensor Spectrum} \label{sec:Inf}


In this section, we explore the conditions under which \eq{V_sugra} can account for
the power spectrum (${\cal P}_\zeta$) and its
scalar spectral index ($n_\zeta$) \cite{Ade:2013zuv},  
\bea
&&{\cal P}_\zeta = (2.198\pm 0.056)\times 10^{-9} \label{powerD} \\
&&n_\zeta = 0.9603\pm 0.0073 \label{spectralD}
\eea
for the first seven e-folds of the inflation after the scale of the last scattering exits the horizon, 
and also the tensor-to-scalar ratio ($r$)  \cite{Ade:2014xna}, 
\dis{
r=0.16^{+0.06}_{-0.05} \label{t/sD}
}
for the first four e-folds.
For the smaller scales, the scalar power spectrum should be smaller than $10^{-4}$.

\begin{figure}
\begin{center}
\subfigure[]
{\includegraphics[width=0.48\linewidth]{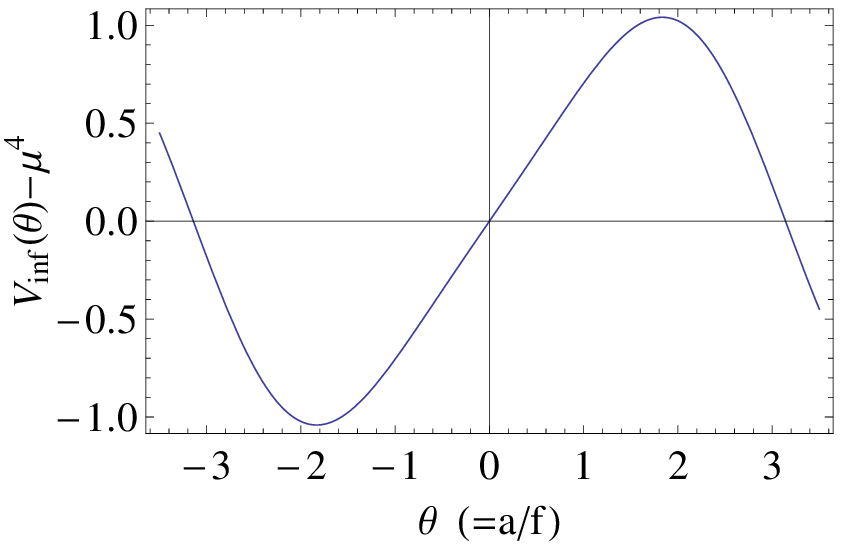}}
\hspace{0.2cm}
\subfigure[] 
{\includegraphics[width=0.48\linewidth]{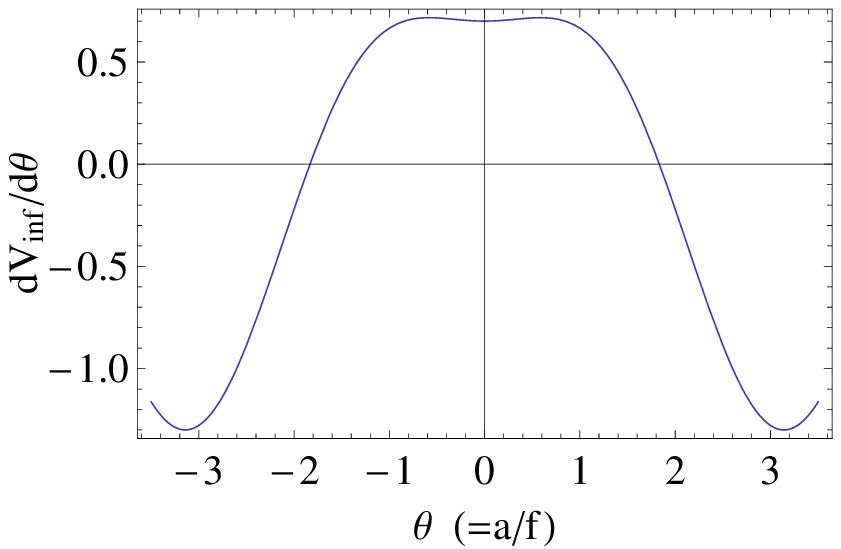}}
\end{center}
\caption{{\bf (a)} Axionic part of the inflaton potential in units of $\mu^4\lambda_1$  
(i.e., ${\rm sin}\theta -\beta {\rm sin}2\theta$)
and   {\bf (b)} $\partial V_{\rm inf}/\partial\theta$ in units of $\mu^4\lambda_1/f$ (i.e. ${\rm cos}\theta -2\beta {\rm cos}2\theta$) with $\beta= 0.15$.   Between $a/f \approx +1$ and $-1$, the slope of the potential is almost constant, and so the $\epsilon_a$ is almost constant. 
  The axionic inflaton $a$ rolls down from $a/f=0.4$ to the minimum given by \eq{amin}. 
Since the potential  becomes  flat around the minimum, the axionic inflaton field slows down and approaches the minimum until the inflation is halted by the waterfall fields.
} \label{fig:Pot}
\end{figure}

These requirements on the primordial spectra can be achieved with the periodic inflaton potential  
by the infinite summation of $e^{-nT/f}$ in \eq{V_sugra}.
However, we will show that all the above cosmological requirements can easily be satisfied with only two terms of the summation.\footnote{It was pointed out that a large running spectral index, 
$dn_\zeta/d{\rm ln}k=-0.028\pm 0.009$ (68\%) \cite{Ade:2014xna} 
can reconcile the tension between BICEP2 and PLANCK on $r$ \cite{runningN}. 
In principle, such a value of $dn_\zeta/d{\rm ln}k$ can also be encoded in the potential \eq{V_sugra} 
with several sinusoidal functions.  
But we need to wait for more data on it at the moment.}  
When all the fields heavier than the Hubble scale are decoupled out, the effective inflationary potential becomes   
\dis{ \label{infPot}
V_{\rm inf}\approx\mu^4\left(1+\lambda_1 {\rm sin}\frac{a}{f}-\lambda_2 {\rm sin}\frac{2a}{f}
+\alpha {\rm log}\frac{\sigma}{\Lambda}\right) ,
}
where the parameters are identified with the parameters in Lagrangian of \eq{V_sugra} as $\mu^4\equiv\kappa^2M^4$, $\lambda_{1,2}\equiv \pm 2im_{1,2}^2/M^2$ ($\ll 1$), and $\alpha\sim \kappa^2/8\pi^2$ ($\ll 1$), respectively.   
Here we have assumed that $m_{1,2}^2$ are purely imaginary. 
The last term corresponds to the quantum correction to the potential generated when the waterfall fields $\psi$ and $\overline{\psi}$ are integrated out: 
$\sigma$ denotes the real component of $S$, and $\Lambda$ is the renormalization scale. 
In this potential, the minimum of the $a$ field is located at 
\dis{
\frac{a_{\rm min}}{f} = -{\rm cos}^{-1} \left[\frac{1-\sqrt{1+32 (\lambda_2/\lambda_1)^2}}{8(\lambda_2/\lambda_1)}\right] , \label{amin}
}
which is slightly smaller than $-\pi/2$ for $|\lambda_2/\lambda_1|\ll 1$. For the shape of $\lambda_{1,2}$ terms in the potential \eq{infPot}, see Fig.~\ref{fig:Pot}, where we set $\lambda_2/\lambda_1=0.15$. 
In this case, $a_{\rm min}/f\approx -1.83$.  
The slow-roll parameters in this model are calculated as follows: 
\dis{
\epsilon_a=&\frac{\Mp^2}{2}\frac{(\mu^8/ f^2) (\lambda_1 {\rm cos}\frac{a}{f}- 2\lambda_2 {\rm cos}\frac{2a}{f} )^2}{V_{\rm inf}^2} \approx 
\frac{\lambda_1\xi^2}{2}\left({\rm cos}\theta-2\beta ~{\rm cos}2\theta\right)^2, \\
\eta_a=&\frac{\Mp^2 (\mu^4/f^2) (- \lambda_1 {\rm sin}\frac{a}{f}+4\lambda_2 {\rm sin}\frac{2a}{f})}{V_{\rm inf}} \approx -\xi^2\left({\rm sin}\theta-4\beta ~{\rm sin}2\theta\right),\\
\epsilon_\sigma=&\frac{\Mp^2}{2} \frac{\alpha^2 \mu^8 /\sigma^2}{V_{\rm inf}^2} \approx
\frac{\alpha}{2\chi^2}, \qquad 
\eta_\sigma= - \frac{\Mp^2 \mu^4 \alpha/\sigma^2}{V_{\rm inf}} \approx
-\frac{1}{\chi^2} ,
}
where the parameters $\xi^2$, $\beta$ and the fields $\theta$, $\chi$ are defined as  
\dis{ \label{definitions}
\xi^2\equiv\frac{M_P^2\lambda_1}{f^2} , 
\quad \beta\equiv \frac{\lambda_2}{\lambda_1} , 
\quad {\rm and} \quad \theta\equiv \frac{a}{f} , 
\quad \chi\equiv \frac{\sigma}{\sqrt{\alpha} M_P} .
} 
Note that $\epsilon_\sigma$ is always relatively suppressed: 
$\epsilon_\sigma\ll|\eta_\sigma|$ for $\alpha\ll 1$.
For $r=0.16$, we need a relatively large $\epsilon$ ($\approx 0.01$).  
As mentioned in the Introduction, $\epsilon_\sigma\approx 0.01$ is hard to get in the SUSY hybrid inflation model:
if $\epsilon_\sigma$ is about $0.01$, 
$\eta_\sigma$ should be much larger than it, violating the slow-roll condition. 
Hence, $\epsilon$ should be dominated by $\epsilon_a$ unlike in the case of Ref.~\cite{NHinf}. 
Since the vacuum energy is almost constant in this model, $\epsilon_a$ should also be kept almost constant for a constant power spectrum during the first seven e-folds. 
This is necessary for consistency with the CMB observations, as mentioned in the Introduction.  
Only two terms of $e^{-nT/f}$ in \eq{V_sugra} would be sufficient, as seen in Fig.~\ref{fig:Pot}-(b).

In the two-field inflationary scenario, the cosmological observables in Eqs.~(\ref{powerD})--(\ref{t/sD}) are expressed 
as follows \cite{GarciaBellido:1995qq,VW,Choi:2007su}:  
\bea
&&{\cal P}_\zeta 
\approx \frac{\mu^4\bar{u}^2}{24\pi^2\Mp^4\epsilon_a^*}\left(1+\hat{r}\right) ,
\label{power} \\
&&n_\zeta -1 \approx -2(\epsilon_a^*+\epsilon_\sigma^*)
+2\frac{-2\epsilon_a^*+\bar{u}^2(\eta_a^*+\eta_\sigma^*\hat{r})}{\bar{u}^2(1+\hat{r})} ,
\label{spectral} \\
&&r
\approx\frac{16\epsilon_a^*}{\bar{u}^2(1+\hat{r})} ,
\label{t/s} 
\eea 
where a (sub- or) superscript ``$*$'' denotes the values evaluated at a few Hubble times after the horizon exit of inflation. 
%
They depend on the effects at the end of inflation as well as the slow-roll parameters at the horizon exit \cite{Lyth:2005qk,Sasaki:2008uc,Byrnes:2008zy,Choi:2012hea}.
The general formula for the power spectrum for two-field inflation including both effects has been obtained in Ref.~\cite{Choi:2012hea}.  
The parameter $\hat{r}$ in the above expressions is defined as the ratio of the contribution from both the axionic and the SUSY hybrid inflatons to the power spectrum, given by
\dis{
\hat{r} = \frac{\epsilon_a^*}{\epsilon_\sigma^*} \frac{\bar{v}^2}{\bar{u}^2},
}
where $\bar{u}$ and $\bar{v}$ are defined as
\dis{
\bar{u}\equiv \frac{R\epsilon_a^e}{\bar{\epsilon}^e} 
\qquad {\rm and}\qquad \bar{v}\equiv \frac{\epsilon_\sigma^e}{\bar{\epsilon}^e} 
} 
with $\bar{\epsilon}^e = R\epsilon_a^e + \epsilon_\sigma^e$, and thus $\bar{u}+\bar{v}=1$.
The (sub- or) superscript $e$'s indicate the values at the end of inflation. 
The factor $R$ parametrizes how much the hypersurface of the two inflaton fields at the end of the inflation deviates from the hypersurface of the uniform energy density,  
%
defined as 
\dis{
R = \frac{\partial_\sigma V_{e}}{\partial_a V_{e}} ~\frac{\partial_a E}{\partial_\sigma E} .
}
Here $V_e$ and $E$ [$=E(a_e,\sigma_e)={\rm constant}$] denote the two inflatons' potential and hypersurface at the end of inflation, respectively.

It turns out that the axionic inflaton $a$ already arrives near the minimum, $a_{}/f\approx-1.80$ at 20 e-folds for $\xi^2=\frac14$ and $\beta=0.15$. 
Inflation is driven dominantly by $\sigma$, following the scenario of the ordinary SUSY hybrid model. 
As $S$ reaches $M$, i.e. $\kappa^2\big||S|^2-M^2\big|\lesssim \kappa^2M^4/M_P^2$, the waterfall fields $\psi$ and $\overline{\psi}$ become light and can start rolling down to their true minima. 
When the slow-roll condition for $S$ is violated due to the nonzero VEVs of $\psi$ and $\overline{\psi}$, $M_P^2\partial^2_{SS^*}V_e/V_e\approx 1$, inflation is eventually over.  
When inflation terminates, the inflatons' potential $V_e$ is given by   
\dis{ \label{endPot}
\frac{V_e}{\kappa^2M^4}\approx& \left|1-\sum_n\frac{m_n^2}{M^2}e^{-nT/f}-\frac{\psi\overline{\psi}}{M^2}\left(1+\sum_n\rho_n e^{-nT/f}\right)\right|^2
\\
&+
\frac{|S|^2}{M^2}\left[\frac{|\psi|^2+|\overline{\psi}|^2}{M^2}\left|1+\sum_n\rho_ne^{-nT/f}\right|^2\right] ,
%
%
}
which is just the scalar potential derived from \eq{globalSUSYPot}. 
Since $\psi$ and $\overline{\psi}$ become light, here we do not consider the logarithmic term of \eq{infPot}.
In \eq{endPot}, we neglected the terms coming from $|\partial W/\partial T|^2$, 
which are of order $m_n^4/(Mf)^2$, $(\psi\overline{\psi})^2/(Mf)^2$, etc., due to the relative smallness.  
%
Since inflation is over with $M_P^2\partial^2_{SS^*}V_e/V_e\approx 1$, it is reasonable to take the slow-roll condition as the hypersurface of the end of inflation, 
$E=M_P^2\partial^2_{SS^*}V_e/V_e$. 
Note that when the slow-roll condition is violated, $\psi$ and $\overline{\psi}$ do not reach their minima yet, i.e. $|\langle\psi\rangle|=|\langle\overline{\psi}\rangle|\sim {\cal O}(10^{14}) \gev$ for $M\sim {\cal O}(10^{16}) \gev$.

In our case, the axionic inflaton field very closely approaches its minimum at the end of inflation, $\partial_a V_{\rm inf} \rightarrow 0$, unlike the case of Ref.~\cite{NHinf}, where $V_{\rm inf}$ is given in \eq{infPot}. 
As a result, $\partial_\sigma V_e$, $\partial_aE$, and also $\partial_aV_e$ are suppressed with ${\cal O}(\psi^2/M^2)$.  
However, $\partial_\sigma E$, which is of order $\psi^4/M^5$, is more suppressed. 
Thus, $R$ is estimated to be of order $M^2/\psi^2\sim 10^4$. 
We suppose that $\rho_n$s in Eqs.~(\ref{globalSUSYPot}) or (\ref{endPot}) are given such that $\partial_aV_e/\partial_\sigma V_e\gtrsim 0.1$ and so $R\epsilon_a^e/\epsilon_\sigma^e\gtrsim 10^2$. 
Then, we have $\bar{u}^2 \approx 1$ and $\bar{v}^2\lesssim 10^{-4}$, which can lead to $\hat{r} \ll 1$ 
only if $\epsilon_a^*$ is not excessively larger than $\epsilon_\sigma^*$ or  $\epsilon_a^*/\epsilon_\sigma^*\lesssim 10^{3}$.
Consequently, the scalar power spectrum can be determined dominantly by the inflaton $a$ under such a proper situation, although the vacuum energy $\mu^4$ results from dynamics of the inflaton $\sigma$ as in the usual SUSY hybrid inflation.

From Eqs.~(\ref{t/sD}) and (\ref{t/s}), $\epsilon_a^*$ should be around $0.01$, 
which determines $\mu=2.08\times 10^{16} \gev$ with Eqs.~(\ref{powerD}) and (\ref{power}) as mentioned in the Introduction.
As seen in Fig.~\ref{fig:Pot}-(b), $\epsilon_a$ can be almost constant between $\theta=a/f\approx +1$ and $-1$ for $\beta=0.15$.  
We take $\theta_*=0.4$ as the initial value of $\theta$. 
Then, both $\epsilon_a^*$ and $\eta_a^*$ can be $0.01$,  
e.g., with $\lambda_1=0.16$ and $\xi^2=\frac14$, determining $f=0.8\Mp$ from \eq{definitions}. 
Assuming $\epsilon_\sigma^*\ll \epsilon_a^*$, thus, we obtain the desired value of $n_\zeta$ ($\approx 0.96$) from \eq{spectral}.
The rolling of the axionic inflaton $a$ from $\theta=0.4$ to $-1.0$ provides eight e-folds:
%
%
\dis{
\Delta N_a\approx\frac{f^2}{M_P^2\lambda_1}\int_{-1.0}^{0.4}\frac{d\theta}{{\rm cos}\theta-2\beta ~{\rm cos}2\theta}
=\frac{1}{\xi^2}\int_{-1.0}^{0.4}\frac{d\theta}{{\rm cos}\theta-2\beta ~{\rm cos}2\theta}
\approx 8 .
}
Therefore, the tensor spectrum is also constant in this $\Delta N_a \approx 8$ with $r\approx 16\epsilon_a^* \approx 0.16$. 
Thus, all the cosmological observables are determined dominantly by dynamics of $a$, 
although inflation is driven by both the inflaton fields, $a$ and $\sigma$, in this period.  
%


After $\Delta N_a\approx 8$, the inflaton $a$ gradually approaches its minimum, but its field value becomes trans-Planckian. 
Analyses on the cosmological observables $\epsilon_a$, $\eta_a$, etc. might loose the predictivity in the super-Planckian regime,
particularly if the shift symmetry is just an accidental symmetry among low-dimensional operators: (higher-dimensional) softly breaking terms could be included in the superpotential (and also the K${\rm \ddot{a}}$hler potential) (See e.g. Ref.~\cite{NaturalChaotic}). 
While Planck-suppressed higher-dimensional operators could be sensitive to trans-Planckian field values, however, they leave intact the above results associated with the sub-Planckian field values.  
We just assume that their effects are small enough 
even in the super-Planckian regime, even if they are not related to the present observational data. 

The second stage of inflation after the first eight e-folds is driven mainly by the SUSY hybrid inflaton field $S$ (or $\sigma$), which maintains until $S$ reaches $M$: i.e., $\sigma_e=\sqrt{2}M$.
Once the waterfall fields $\psi$ and $\overline{\psi}$ become light, 
the potential of $a$ is dominated by the $\rho_n$ terms as discussed in \eq{endPot}. 
During the whole period of inflation, an e-folding number of $50$ is obtained by $\sigma$: 
\dis{
N_\sigma=\frac{1}{\Mp^2}\int_e^* \frac{V}{\partial V/\partial \sigma}d\sigma \approx  \frac{1}{M_P^2}\int_e^*\frac{\sigma}{\alpha}d\sigma
=\frac12\left(\chi_*^2-\chi_e^2\right)=50  , 
}
so we have $\chi_*^2=\sigma_*^2/\alpha M_P^2\approx 100$. 
$\sigma$ is always required to be sub-Planckian,  $\sigma < \Mp$, which thus yields
\dis{
\frac{\sigma_*}{M_P} \approx 10\sqrt{\alpha} ~ < ~ 1  \quad {\rm or} \quad 
\kappa ~ < ~ 0.9 .
}
For $\alpha\ll 1$, hence, $\epsilon_\sigma^*=\alpha/2\chi_*^2\ll 0.01$, 
which meets the previous assumption, $\epsilon_\sigma^*\ll \epsilon_a^*$.  
Since $\mu$ is just the GUT scale, $M_G$, $M$ ($=\mu/\sqrt{\kappa}$) is slightly higher than the GUT scale. 
For $\sigma_*/M_P\approx 0.3$ ($1.0$), we find $\kappa\approx 0.3$ ($0.9$) and $M\approx 1.9\times M_G$ ($1.0\times M_G$).   
They yield $\epsilon_\sigma^*\sim 10^{-5}$ ($10^{-4}$), which maintains the consistency with $\hat{r}\ll 1$.

\section{Conclusion} \label{sec:conclusion}

The large tensor-to-scalar ratio, $r\approx 0.16$ observed by the BICEP2 Collaboration requires 
a super-Planckian field variation during the inflationary era in the single-field inflation scenario, 
which might destroy field theory description on inflation. 
Although it could be controlled using a shift symmetry, a super-Planckian decay constant would be another obstacle for constructing an inflation model based on effective field theory.

In this paper, we have considered a two-field inflationary scenario 
based on the NH inflationary model, 
which combines the natural and SUSY hybrid inflation models. 
In this model, the Hubble scale induced mass term for the two inflaton fields, $\sigma$ and $a$ 
can be avoided by introducing the U(1)$_R$ and a shift symmetries, and employing the minimal form of the K${\rm \ddot{a}}$hler potential. 
The inflation is terminated by the  VEVs of the waterfall fields, which is around the GUT scale. 

The power spectrum in the observable scale of $\Delta N \sim 8$ is determined mainly by the axionic inflaton field, while the needed e-folding number can be obtained from the hybrid inflation sector 
before the waterfall field terminates inflation.
We find that the large tensor spectrum corresponding to $r\approx 0.16$ 
as well as the spectral index of the scalar power spectrum $n_\zeta\approx 0.96$ 
are well obtained 
with a sub-Planckian decay constant, $f\lesssim M_P$. 
While the field values of the SUSY hybrid inflaton $\sigma$ are sub-Planckian ($\sim 10^{16-17} \gev$) during inflation, 
the axionic inflaton $a$ moves a Planckian distance along the phase. 
However, the cosmological observables are determined while it stays in the sub-Planckian regime.

\acknowledgments

\noindent 
K.-Y.C. appreciates Asia Pacific Center for Theoretical Physics for its support of the Topical Research Program.
This research is supported by Basic Science Research Program through the 
National Research Foundation of Korea (NRF) funded by the Ministry of Education, 
Grants No. 2011-0011083 (K.-Y.C.) and  No. 2013R1A1A2006904 (B.K.). 
B.K. acknowledges the partial support 
by Korea Institute for Advanced Study (KIAS) grant funded by the Korean government.


\end{document}